\documentclass{eptcs}

\usepackage{amssymb}
\usepackage{amsmath}
\usepackage{stmaryrd}
\usepackage[all]{xy}
\usepackage{tikz}

\title
{The Boolean Algebra of Cubical Areas as a Tensor Product in the Category of
  Semilattices with Zero}

\author{Nicolas Ninin
\institute{CEA, LIST and University Paris-Sud, France}
\email{nicolas.ninin@cea.fr}
\and
Emmanuel Haucourt
\institute{CEA, LIST}
\email{emmanuel.haucourt@cea.fr}
}
\usepackage{our_macros}

\def\titlerunning{Cubical areas as tensor product}
\def\authorrunning{ N. Ninin and E. Haucourt }

\newtheorem{thm}{Theorem}[section]
\newtheorem{lem}{Lemma}[section]
\newtheorem{prop}{Proposition}[section]

\newtheorem{cor}{Corollary}[section]

\begin{document}

\maketitle

\begin{abstract}
In this paper we describe a model of concurrency enjoying an algebraic
structure reflecting the parallel composition.
For the sake of simplicity we restrict to linear concurrent programs
i.e. the ones with neither loops nor branchings.
Such programs are given a semantics using cubical areas that we call
geometric.
The collection of all cubical areas admits a structure of
tensor product in the category of semi-lattice with zero.
These results naturally extend to fully fledged concurrent programs up
to some technical tricks.
\end{abstract}

\section{Introduction}
In the two last decades, many geometrical or topological models of
concurrent programs have emerged
\cite{Dijkstra68,CR87,Pratt91,Pratt00,Grandis03,FGR06,Krishnan09}. We
are especially interested in a simple geometrical
one based on the so-called  $n$-dimensional cubical areas which model
the control flow for parallel composition of threads without loops or 
branchings.
They actually
form a boolean algebra $\B_{\R^n}$ whose operations are pervasively used in 
\cite{BH10}. The purpose of our paper is to formalize the fact that these
operations are actually deduced from their analog in
$\B_{\R}$.
Formally we prove that the tensor product of two boolean algebras is still a
boolean algebra when it is considered in the category of semilattice with zero
($\slatz$). We then show that the boolean algebra  $\B_{\R^n}$, which is in
particular a semilattice with zero, can be seen as such a product.

The class of concurrent program we study arises from a toy language
manipulating mutex. Using Djikstra's notation \cite{Dijkstra68}, we
consider processes to be sequences of locking operations $Pa$ on
mutex $a$ and unlocking operations $Va$.
To each concurrent program made of $n$ processes we have a subset of
$\R^n$ representing its consistent states. By construction, such subsets of $\R^n$ are
finite union of $n$-cubes. They are called cubical areas. The points of these
subsets are to be considered as the states of the PV program. Holes in these subsets
arise from synchronizations between processes. The set of increasing paths on them
then overapproximate the collection of execution traces, and we have a natural
equivalence relation upon increasing paths such that equivalent paths have the
same effect over the system \cite{FGR06}.

\begin{figure}
\begin{center}
\framebox{
\begin{tikzpicture}[thick,scale=0.6]
\draw (1,-0.5) node {Pa};
\draw (4,-0.5) node {Va};
\draw (-0.5,2) node {Pa};
\draw (-0.5,3) node {Va};
\draw (-0.3000,0.0000)--(5.0000,0.0000);
\draw (0.0000,-0.3000)--(0.0000,5.0000);
\draw (4.0000,2.0000)--(4.0000,3.0000);
\draw (1.0000,2.0000)--(4.0000,2.0000);
\draw (1.0000,2.0000)--(1.0000,3.0000);
\draw (1.0000,3.0000)--(4.0000,3.0000);
\fill[lightgray] (1.0000,2.0000)--(1.0000,3.0000)--(4.0000,3.0000)--(4.0000,2.0000)--cycle;

\begin{scope}[xshift=200]

\draw (2,-0.5) node {Pb};
\draw (3,-0.5) node {Vb};
\draw (-0.5,1) node {Pb};
\draw (-0.5,4) node {Vb};

\draw (-0.3000,0.0000)--(5.0000,0.0000);
\draw (0.0000,-0.3000)--(0.0000,5.0000);
\draw (2,1)--(2,4)--(3,4)--(3,1);

\fill[lightgray] (2.0000,1.0000)--(2.0000,4.0000)--(3.0000,4.0000)--(3.0000,1.0000)--cycle;
\end{scope}
\begin{scope}[xshift=400]
\draw (1,-0.5) node {Pa};
\draw (2,-0.5) node {Pb};
\draw (3,-0.5) node {Vb};
\draw (4,-0.5) node {Va};

\draw (-0.5,1) node {Pb};
\draw (-0.5,2) node {Pa};
\draw (-0.5,3) node {Va};
\draw (-0.5,4) node {Vb};
\draw (-0.3000,0.0000)--(5.0000,0.0000);
\draw (0.0000,-0.3000)--(0.0000,5.0000);

\fill[lightgray] (1.0000,2.0000)--(1.0000,3.0000)--(4.0000,3.0000)--(4.0000,2.0000)--cycle;

\fill[lightgray] (2.0000,1.0000)--(2.0000,4.0000)--(3.0000,4.0000)--(3.0000,1.0000)--cycle;
  \draw (3.0000,3.0000)--(4.0000,3.0000);
\draw (3.0000,2.0000)--(4.0000,2.0000);
\draw (4.0000,2.0000)--(4.0000,3.0000);
\draw (2.0000,4.0000)--(3.0000,4.0000);
\draw (2.0000,1.0000)--(3.0000,1.0000);
\draw (3.0000,3.0000)--(3.0000,4.0000);
\draw (3.0000,1.0000)--(3.0000,2.0000);
\draw (1.0000,3.0000)--(2.0000,3.0000);
\draw (1.0000,2.0000)--(2.0000,2.0000);
\draw (2.0000,3.0000)--(2.0000,4.0000);
\draw (2.0000,1.0000)--(2.0000,2.0000);
\draw (1.0000,2.0000)--(1.0000,3.0000);
\end{scope}
  \end{tikzpicture}}
  \end{center}
\caption{The Swiss flag; At the left the forbiden region of mutex a,
  at the center the forbiden region of mutex b, and the union of the
  two.}
  \label{fig:swiss_flag}
\end{figure}
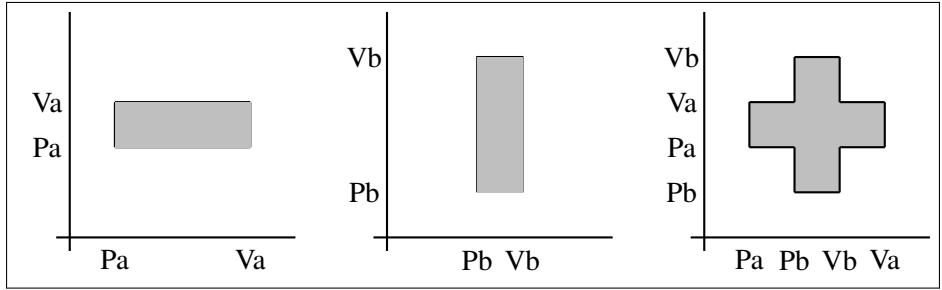

We provide a motivating example for the result to be developped in the paper.
Consider the following program, written in PV language \cite{Dijkstra68},
that consists of two parallel processes $T_1=Pa.Pb.Vb.Va$ and $T_2= Pb.Pa.Va.Vb$
where $a$ and are mutex. Any PV program can be given a geometric semantics
\cite{CR87}, in our specific example it boils down to the so-called ``Swiss
flag'', Fig.~\ref{fig:swiss_flag}, regarded as a subset of $\R^2$.
The (interior of the) horizontal rectangle comprises global states
that are such that $T1$ and $T2$ both hold a lock on $a$, which is not allowed
by the very definition of a mutex. Similarly, the (interior of the)
vertical rectangle consists of states violating the mutual exclusion
property on $b$. Therefore both rectangles form the set of inconsistent states,
which is the complement of $\intp{T_1|T_2}$ the
cubical area of (consistent) states \ie the model of the program. A cubical
area (of dimension $n$) is a finite union of $n$-dimensional parallelepipeds (or
$n$-cubes for short) \ie $n$-fold cartesian products of intervals of $\R$. All
geometric models of PV programs actually arise as cubical areas whose dimension
is the number of processes the program is made of. More precisely the algorithm
producing the geometric model of a PV program first returns the cubical area of
its inconsistent states and then computes the set theoretic complement of the
later to obtain the actual model of the program. For example the \emph
{deadlock attractor} of the program \ie the subset of points of the
geometric model from which all emerging paths can be extended to a path
ending at a deadlock, is also a cubical area. The collection of
$n$-dimensional cubical areas indeed forms a boolean subalgebra of the powerset
$2^{\R^n}$. Moreover the cubical areas can be handled automatically which makes them
suitable for implementation, this practical fact is at the origin of our interest
for them. It is also worth to notice that the boolean algebra of cubical sets
actually provides the ground upon which the static analyzer ALCOOL is based.
The following property of the geometric semantics of the PV language is also crucial: 
suppose we are given two groups of processes $P_1,\ldots,P_n$ and
$Q_1,\ldots,Q_m$ so their sets of occuring resources are disjoint, then
\[
\intp{P_1|\cdots|P_n|Q_1|\cdots|Q_m} =
\intp{P_1|\cdots|P_n} \times \intp{Q_1|\cdots|Q_m}
\]
from which one can (rather easily) deduce that
\[
\B_{\intp{P_1|\cdots|P_n|Q_1|\cdots|Q_m}} =
\B_{\intp{P_1|\cdots|P_n}} \otimes \B_{\intp{Q_1|\cdots|Q_m}}
\]
where $\B_{\intp{X}}$ denotes the boolean algebra of subareas of the model
$\intp{X}$ of a PV program $X$. Conversely one may ask whether a
tensor decomposition of $\B_{\intp{X}}$ indicates a potential parallelization
of $X$ \ie gathering its processes in groups that do not interact with each
other; and even more theoretically whether  $\B_{\intp{X}}$ admits a
prime decomposition \cite{BH10}. The purpose of this paper is to define and
study the aforementioned tensor product.
%

First remark that $1$-dimensional cubical areas are the finite unions of
intervals of the real line. Our main goal is then to prove that the boolean algebra
of $n$-dimensional cubical areas is the $n$-fold tensor product of the
boolean algebra of $1$-dimensional cubical areas.
The main obstacle resides in the bad behavior of the tensor product in the category of boolean algebras which is actually
degenerated. Yet we have finally discovered that the category of semilattices with zero is the right framework for our
purpose. It is worth to notice that the
zero hypothesis (the presence of a least element) cannot be dropped.
\\ \\
\textbf{Outline of the paper.}\\
Section 2 defines cubical areas, and provides details about
their boolean structure. Section 3 introduces the
notion of tensor product in a category, and shows that the tensor
product of two boolean algebras in $\slatz$ is still a boolean
algebra. Section 4 relates the boolean algebra of cubical areas to the tensor product
by proving $\B_{\R}\otimes\B_{\R}\simeq \B_{\R^2}$.
\section{Cubical Area}
A \emph{cube} of dimension $n\in\N$ (or just $n$-\emph{cube}) is the set
product of a $n$-uple of (potentially unbounded) intervals of the real line $\R$. It is therefore
a subset of $\R^n$. A \emph{maximal subcube} of $X\subseteq\R^n$ is a
cube $C\subseteq X$ such that $C=C'$ holds for all cubes
$C'$ such that $C \subseteq C' \subseteq X$. The union of any $\subseteq$-chain
of $n$-cubes is a cube. As a consequence any subcube of $X$ is
contained in a maximal subcube of $X$. A \emph{cubical cover} of $X$ is a
family of cubes whose union is $X$. Then we define $\alpha(X)$ as the
collection of all maximal subcubes of $X$. Given $\C$ and $\C'$ two
families of $n$-cubes we define $\gamma(\C)$ as the union of all the elements
of $\C$ and write $\C \preccurlyeq \C'$ when any element of $\C$ is
contained in some element of $\C'$. We call a \emph{cubical area} any
subset of $\R^n$ admitting a finite cubical cover.

\textbf{Example of a cubical area of $\R^2$}

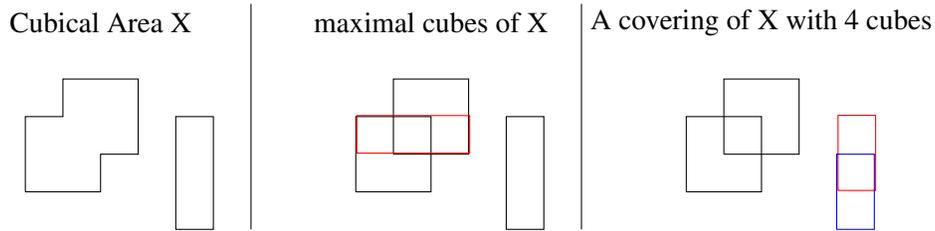
\begin{center}




\begin{tikzpicture}[scale=0.5]

\draw (2,5.5) node {Cubical Area X};
\draw (0,1) -- (0,3) -- (1,3) -- (1,4) -- (3,4) -- (3,2) -- (2,2) -- (2,1) -- (0,1) ;
\draw (4,0) -- (4,3) -- (5,3) -- (5,0) -- (4,0) ;
\draw (6,0) -- (6,6) ;

\begin{scope}[xshift=250]
\draw (2,5.5) node {maximal cubes of X};
\draw (0,1) -- (0,3) -- (2,3) -- (2,1) -- (0,1);
\draw (1,2) -- (1,4) -- (3,4) -- (3,2) -- (1,2);
\begin{scope}[xshift=0.8, yshift=0.8]
\draw [color=red] (0,2) -- (0,3) -- (3,3) -- (3,2) -- (0,2);
\end{scope}
\draw (4,0) -- (4,3) -- (5,3) -- (5,0) -- (4,0) ;
\draw (6,0) -- (6,6) ;
\end{scope}

\begin{scope}[xshift=500]
\draw (2,5.5) node {A covering of X with 4 cubes };
\draw (0,1) -- (0,3) -- (2,3) -- (2,1) -- (0,1);
\draw (1,2) -- (1,4) -- (3,4) -- (3,2) -- (1,2);
\begin{scope}[xshift=1, yshift=1]
\draw [color=red] (4,1) -- (4,3) -- (5,3) -- (5,1) -- (4,1) ;
\end{scope}
\draw [color=blue] (4,0) -- (4,2) -- (5,2) -- (5,0) -- (4,0) ;
\end{scope}

\end{tikzpicture}
\end{center}
%
\begin{lem}
\label{intersection of cubical families}
Let $\C$ and $\C'$ be families of $n$-cubes that contain all the maximal
subcubes of their unions $\gamma(\C)$ and $\gamma(\C')$. Then the family of $n$-cubes
\[
\setof{C \cap C'}{C\in\C\text{ and }C'\in\C'}
\]
contains all the maximal subcubes of $\gamma(\C) \cap \gamma(\C')$.
\end{lem}
Let $C''$ be a subcube of $\gamma(\C) \cap \gamma(\C')$ and let $C$ and $C'$ be subcubes of $\gamma(\C)$
and $\gamma(\C')$ respectively such that $C''\subseteq C$ and $C''\subseteq C'$. Then $C\cap C'$
is a subcube of $\gamma(\C)\cap \gamma(\C')$ containing $C''$.
%
\begin{lem}
\label{complement of cube}
The complement of any $n$-cube admits at most $2n$ maximal subcubes
\end{lem}
Let $I_1 \times \cdots \times I_n$ be a cube, then any maximal subcube of its complement can be written
as
\[
\R\times\cdots\times\hspace{-6mm}\underbrace{J_k}_{k\text{th position}}\hspace{-6mm}\times\cdots\times\R
\]
with $J_k$ being a maximal subinterval of the complement of $I_k$ in $\R$.
%
Given $X\subseteq\R^n$ we denote the complement of $X$ in $\R^n$ by $X^c$.
\begin{prop}
\label{cns for cubical areas}
A subset of $\R^n$ is a cubical area iff it has finitely many maximal subcubes.
\end{prop}
%
\begin{cor}
\label{boolean algebra of cubical areas}
The collection $\B_{\R^n}$ of all the $n$-cubical areas is a boolean subalgebra of
the powerset of $\R^n$.
\end{cor}
The empty set and $\R^n$ are cubical areas. From what we have seen before it
is clear that $\B_{\R^n}$ is stable under complement and binary
intersection. From De Morgan laws it is also stable under binary
unions.
%
\section{Tensor Product of Boolean Algebra}
%
Tensor products of vector spaces are  well-known,
but they exist in many other categories provided with a forgetful
functor to $\Set$ \cite {Borceux94b}. The categories of boolean algebra, 
distributive lattices, and semilattices with zero are examples of such structures.

Given three objects $A$, $B$ and
$X$  of the same category, a \emph{bimorphism} from $A,B$ to $X$
is a set map $f:A\times B\to X$ such that for all $a\in A$ and for all
$b\in B$ the mappings $f(a,\_):B\to X$ and $f(\_\ ,b):A\to X$ are
morphisms.
Given a bimorphism $i:A\times B \to X$ we say that $X$ is a \emph{tensor 
product} of $A$ and $B$ if for every object $C$ and every bimorphism 
$f:A\times B \to C$ there exists a unique morphism $h:X\to C$ such that $f=h\circ i$. 
Tensor products are unique up to isomorphism and they are denoted by $A\otimes B$. 
The bimorphism $i$ is not surjective yet its image generates $A\otimes B$. 
In particular $A\times B$ is a subset of $A\otimes B$ whose elements are said 
to be generating, we write $i(a,b)=a\otimes b$.\\

\textbf{Example of a bimorphism in $\R^2$}\\
Let $f:\B_{\R}\times \B_{\R}\rightarrow C$ be a bimorphism in
$\slatz$. An element of $\B_{R}$ is just a union of segments which may be either
open, close, or both (ie a cubical area) of $\R$. By definition of a bimorphism
we have $f(0_{\B_{\R}},b)=f(a,0_{\B_{\R}})=0_C$, where $a,b \in
\B_{R}$ and $0_{\B_{\R}}$ is the empty set (of $\R$), and also
$f(a_1\cup_{\B_{\R}} a_2,b)=f(a_1,b)\cup_C f(a_2,b)$. Consider for example
$a_1=]0,1]$,$a_2=[1,2]$, $b_1=[0,1]$,$b_2=[1,2]$, let $a=a_1\cup
a_2=[0,2]=b$, then it comes \[f(a_1,b_2)\cup f(a_2,b_2)\cup
f(a,b_1)=f(a_1\cup a_2,b_2)\cup f(a_2,b_2)=(a_1\cup a_2, b_1\cup
b_2)=f(a,b)\]  It geometrically means that $f$ is constant on the
cubical area $[0,2]^2$, even if you subdivide it.

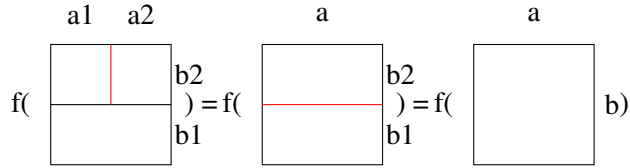
\begin{center}
\begin{tikzpicture}[scale=0.8]
\draw (0.5,2.5) node {a1};
\draw (1.5,2.5) node {a2};
\draw (2.3,0.5) node {b1};
\draw (2.3,1.5) node {b2};

\draw (-0.5,1)node {f(};

\draw (0,0) -- (0,2) -- (2,2) -- (2,0) -- (0,0) ;
\draw (0,1) -- (2,1) ;
\draw [color=red](1,2) -- (1,1) ;
\draw (2.5,1)node {) =};

\begin{scope}[xshift=100]
\draw (1,2.5) node {a};
\draw (2.3,0.5) node {b1};
\draw (2.3,1.5) node {b2};
\draw (-0.5,1)node { f(};
\draw (0,0) -- (0,2) -- (2,2) -- (2,0) -- (0,0) ;
\draw [color=red](0,1) -- (2,1) ;
\draw (2.5,1)node {) =};
\end{scope}

\begin{scope}[xshift=200]
\draw (1,2.5) node {a};
\draw (2.3,1) node {b};
\draw (-0.5,1)node {f(};
\draw (0,0) -- (0,2) -- (2,2) -- (2,0) -- (0,0) ;
\draw (2.5,1)node {)};
\end{scope}
\end{tikzpicture}

\end{center}
 \vspace{1em}
Formally speaking, a \emph{boolean algebra} is a distributive lattice together with an involution, the so-called \emph{complement}, $x\in X\mapsto x^c\in X$ satisfying
$x \vee x^c = 0$ and $x \wedge x^c = 1$ for all $x\in X$, where $0$ and $1$
are the neutral elements for $\vee$ and $\wedge$ respectively.
In particular any boolean algebra is also a
bounded distributive lattice, a semi-lattice with zero
etc, and all of these structures induce its own tensor product. Among the corresponding
categories we look for the one in which the $n$-fold tensor product of
$\B_{\R}$ is isomorphic with $\B_{\R^n}$. As we shall see, this isomorphism is
actually an isomorphism of boolean algebras.

%
For example let $f$ be a bimorphism
of bounded lattices from $A$, $B$ to $X$; given $a\in A$ and $b\in B$
we have $f(0_A,b)=0_X$ and $f(a,1_B)=1_X$, thus $0_X=f(0_A,1_B)=1_X$. Hence
the set of bimorphism from $A\times B\rightarrow X$ is a singleton if $X$ is degenerated; empty otherwise. In
other words $A\otimes B$ is degenerated. For similar reasons the tensor product
in $\Bool$ (\resp in bounded lattice or distributive bounded lattice)
is irrelevant. Indeed we ultimately want to recover $\B_{\R^n}$ from $\B_{\R}$.

Tensor products of semilattices and related structures have already been
the subject of many publications \cite
{AK78,Fraser76a,Fraser78,Shmuely79,GLQ81,GW00}. In particular the next
theorem has been proved in \cite{Fraser76b} for semilattices. Minor
changes lead to the result for semilattices with zero.

\begin{thm}
\label{slatz}
The collection of distributive lattices with zero is stable under finitary
tensor product in $\slatz$.
Moreover given distributive lattices $A,B$, and $a_i$, $b_i$
elements of $A$ and $B$ respectively, we have:
\[(a_1\otimes b_1) \wedge (a_2\otimes b_2)=(a_1\wedge a_2)\otimes (b_1\wedge b_2)\]

\end{thm}
From now, unless otherwise stated, all the tensor products are understood
in $\slatz$.
\begin{prop}
\label{tensor product of boolean algebras}\ \\
The tensor product (in $\slatz$) of a pair of boolean algebras is a boolean
algebra
\end{prop}
The previous theorem gives us solid ground to prove the
proposition. A boolean algebra being a distributive lattice
with complement, it suffices to find a candidate for $(a\otimes b)^c$ for every element $a \in A$,
$b \in B$ with $A,B \in \Bool$.
\begin{lem}
Given a pair of boolean algebras $A$, $B$ and $a\in A$, $b\in B$ we have:
 \[   \left(a\otimes b\right) \vee \left(\left(1_A\otimes b^c\right
  )\vee \left(a^c\otimes 1_B\right)\right)=1 \;\ \ \ \ and\ \ \ \ \;
    \left(a\otimes b\right) \wedge \left(\left(1_A\otimes b^c\right
  )\vee \left(a^c\otimes 1_B\right)\right)=0 \]
\end{lem}
\textbf{proof.} First we need to expand the $1$ either as $a\vee a^c$ or $b\vee b^c$
\[  (a\otimes b) \vee \big((1_A\otimes b^c)\vee (a^c\otimes 1_B)\big)= (a\otimes b) \vee \big((a\vee a^c)\otimes b^c)\vee (a^c\otimes (b\vee b^c)\big)
\]
Using the fact that $(a\otimes b)\vee
(a\otimes c)=a\otimes(b\vee c)$ and that $a\vee a^c=1$, we expand and reduce to obtain $1$.
The second equality is obtained the same way distributing $\vee$ over $\wedge$
\[
(a\otimes b) \wedge \big((1_A\otimes b^c)\vee (a^c\otimes
1_B)\big)=(a\otimes b) \wedge (1_A\otimes  b^c) \vee (a\otimes
b)\wedge (a^c\otimes 1_B)
\]
Similarly we prove that the preceding expression reduces to $0$. \hfill $\square$\\

Every generating element (\ie of the form $a\otimes b$) thus has a complement, and 
any element is a finite union of generating elements
$x=\bigvee_{i\in I} (a_i\otimes b_i)$.
The existence of a complement then follows from the De Morgan's law:
\[((a_1\otimes b_1)\vee (a_2\otimes b_2))^c=(a_1\otimes b_1)^c\wedge
(a_2\otimes b_2)^c\]
The later essentially derives from the relation, $(a_1\otimes b_1) \wedge
(a_2\otimes b_2)=(a_1\wedge a_2)\otimes(b_1\wedge b_2)$ which is provided by
Theorem \ref{slatz}.
\section{The collection of cubical areas $\B_{\R\times\R}$ as a tensor product}
\begin{thm}
\label{cubical areas are tensor products}
The tensor product $\B_\R \otimes \B_\R$ in $\slatz$ is actually a boolean
algebra that is isomorphic (as boolean algebras) with $\B_{\R\times\R}$.
\end{thm}
We prove that $\B_{\R\times\R}$ satisfies the universal property 
characterizing the tensor product.
Let $X\in\slatz$ and $f:\B_{\R}\times  \B_{\R}\rightarrow X$ be a
bimorphism in $\slatz$.
We want to find a morphism $h:\B_{\R\times\R}\rightarrow X$ such that
the diagram commutes :
\[
\xymatrix@R4mm{
 \B_{\R}\times  \B_{\R}\ar[dr]_f \ar@{^{(}->}[r]^i & \B_{\R\times\R} \ar@{-->}[d]^h
  \\
                                  & X &   }
\]
where $i$ is the canonical inclusion. We define $h$ on the image of $i$ by
$h(i(I_1,I_2))=f(I_1,I_2)$ with $I_1,I_2\in \B_{\R}$. Since $h$ has to be a morphism this
definition extends to all $\B_{\R\times\R}$ with $h(C_1\cup
C_2)=h(C_1)\vee h(C_2)$ where the $C_i$'s are generating elements of
$\B_{\R\times\R}$ \ie elementary cubes which we write $a\times b$.
This mapping might however not be well defined since a cubical area of $\R^2$
can be covered by smaller cubes in infinitely many ways. So it remains to check
the soundness of the definition.

\begin{lem}
 Let $h$ be defined as above, and let $X=\bigcup_{i\in I}C_i= \bigcup_{j\in
   J}C'_j$ be a cubical area described as two finite unions of
 generating elements $C_i$ and $C_j$, then
\[ \bigvee_{i\in I}h(C_i)= \bigvee_{j\in J}h(C'_j)\]
and thus $h$ is well defined .
\end{lem}
\textbf{Example in $\R^2$:}\\
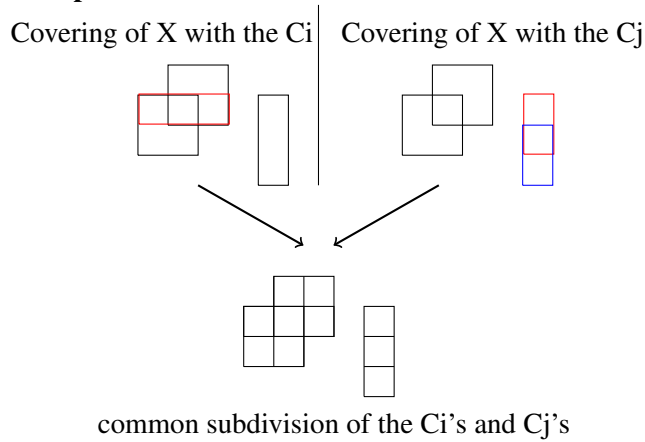
\begin{minipage}{0.55\textwidth}

\begin{tikzpicture}[scale=0.4]

\draw (0.8,5) node {Covering of X with the Ci};
\draw (0,1) -- (0,3) -- (2,3) -- (2,1) -- (0,1);
\draw (1,2) -- (1,4) -- (3,4) -- (3,2) -- (1,2);
\begin{scope}[xshift=1, yshift=1]
\draw [color=red] (0,2) -- (0,3) -- (3,3) -- (3,2) -- (0,2);
\end{scope}
\draw (4,0) -- (4,3) -- (5,3) -- (5,0) -- (4,0) ;
\draw (6,0) -- (6,6) ;

\begin{scope}[xshift=250]
\draw (3,5) node {Covering of X with the Cj };
\draw (0,1) -- (0,3) -- (2,3) -- (2,1) -- (0,1);
\draw (1,2) -- (1,4) -- (3,4) -- (3,2) -- (1,2);
\begin{scope}[xshift=1, yshift=1]
\draw [color=red] (4,1) -- (4,3) -- (5,3) -- (5,1) -- (4,1) ;
\end{scope}
\draw [color=blue] (4,0) -- (4,2) -- (5,2) -- (5,0) -- (4,0) ;
\end{scope}

\draw [thick,->] (2,0) -- (5.5,-2) ;
\draw [thick,->] (10,0) -- (6.5,-2) ;

\begin{scope}[xshift=100, yshift=-200]
\draw (3,-1) node {common subdivision of the Ci's and Cj's};
\draw (0,1) -- (0,3) -- (2,3) -- (2,1) -- (0,1);
\draw (1,2) -- (1,4) -- (3,4) -- (3,2) -- (1,2);
\draw  (0,2) -- (0,3) -- (3,3) -- (3,2) -- (0,2);
\draw (4,0) -- (4,3) -- (5,3) -- (5,0) -- (4,0) ;
\draw (1,1) -- (1,4);
\draw (2,1) -- (2,4);
\draw (4,1) -- (5,1);
\draw (4,2) -- (5,2);
\end{scope}
\end{tikzpicture}
\end{minipage}\hfill
\begin{minipage}{0.43\textwidth }
Consider the first cubical area $X$ met in section 2. We can find a
common subdivision of the $C_i$ and the $C_j$, by cutting along
every hyperplane supporting an edge of a cube. We know that $h(a\otimes b)=f(a,b)$ for any generating
element. Since $f$ is a
bimorphism we can glue two cubes sharing a face. 
By induction we get that the value of
$h$ is the same on those three families of cubes.
\end{minipage}
 \\ \\
\textbf{Perspectives.}

These results extend to cartesian products of geometric realizations of
graphs (instead of $\R^n$) so one can take programs with branchings and
loops into account. It means that we can substitute in this paper, connected subsets of
the geometric realization of a graph to the intervals of $\R$. The
graphs of interest being the control flow graphs of threads \cite{Allen70}.
%

%
%
%
%
%
%
%
\bibliographystyle{eptcs}
\bibliography{my_usual_bibliography}
%
%
%
\end{document}